\documentstyle[epsfig,twocolumn,aps,prd]{revtex}
\def \b{B^0}
\def \bea{\begin{eqnarray}}
\def \beq{\begin{equation}}
\def \d{D^0}
\def \eea{\end{eqnarray}}
\def \eeq{\end{equation}}
\def \efi{Enrico Fermi Institute Report No. EFI}
\def \ipp{I_{\pi \pi}}
\def \k{K^0}
\def \m{{\cal M}}
\def \ob{\overline{B^0}}
\def \od{\overline{D^0}}
\def \ok{\overline{K}^0}
\def \s{\sqrt{2}}
\def \tl{\tilde{\lambda}}
\def \ajp#1#2#3{Am.\ J. Phys.\ {\bf#1}, #2 (#3)}
\def \apny#1#2#3{Ann.\ Phys.\ (N.Y.) {\bf#1}, #2 (#3)}

\def \cn{Collaboration}
\def \cp89{{\it CP Violation,} edited by C. Jarlskog (World Scientific,
Singapore, 1989)}
\def \econf#1#2#3{Electronic Conference Proceedings {\bf#1}, #2 (#3)}
\def \epjc#1#2#3{Eur.\ Phys.\ J. C {\bf#1}, #2 (#3)}
\def \f79{{\it Proceedings of the 1979 International Symposium on Lepton and
Photon Interactions at High Energies,} Fermilab, August 23-29, 1979, ed. by
T. B. W. Kirk and H. D. I. Abarbanel (Fermi National Accelerator Laboratory,
Batavia, IL, 1979}
\def \hb87{{\it Proceeding of the 1987 International Symposium on Lepton and
Photon Interactions at High Energies,} Hamburg, 1987, ed. by W. Bartel
and R. R\"uckl (Nucl.\ Phys.\ B, Proc.\ Suppl.\, vol. 3) (North-Holland,
Amsterdam, 1988)}

\def \ibj#1#2#3{~{\bf#1}, #2 (#3)}
\def \ichep72{{\it Proceedings of the XVI International Conference on High
Energy Physics}, Chicago and Batavia, Illinois, Sept. 6 -- 13, 1972,
edited by J. D. Jackson, A. Roberts, and R. Donaldson (Fermilab, Batavia,
IL, 1972)}
\def \ijmpa#1#2#3{Int.\ J.\ Mod.\ Phys.\ A {\bf#1}, #2 (#3)}
\def \ite{{\it et al.}}
\def \jhep#1#2#3{JHEP {\bf#1}, #2 (#3)}

\def \lg{{\it Proceedings of the XIXth International Symposium on
Lepton and Photon Interactions,} Stanford, California, August 9--14 1999,
edited by J. Jaros and M. Peskin (World Scientific, Singapore, 2000)}
\def \lkl87{{\it Selected Topics in Electroweak Interactions} (Proceedings of
the Second Lake Louise Institute on New Frontiers in Particle Physics, 15 --
21 February, 1987), edited by J. M. Cameron \ite~(World Scientific, Singapore,
1987)}
\def \kdvs#1#2#3{Kong.\ Danske Vid.\ Selsk., Matt-fys.\ Medd.\ {\bf #1}, No.~#2
(#3)}
\def \ky85{{\it Proceedings of the International Symposium on Lepton and
Photon Interactions at High Energy,} Kyoto, Aug.~19-24, 1985, edited by M.
Konuma and K. Takahashi (Kyoto Univ., Kyoto, 1985)}

\def \nat#1#2#3{Nature {\bf#1}, #2 (#3)}
\def \nc#1#2#3{Nuovo Cim.\ {\bf#1}, #2 (#3)}
\def \np#1#2#3{Nucl.\ Phys.\ {\bf#1}, #2 (#3)}
\def \npbps#1#2#3{Nucl.\ Phys.\ B Proc.\ Suppl.\ {\bf#1}, #2 (#3)}
\def \os{XXX International Conference on High Energy Physics, Osaka, Japan,
July 27 -- August 2, 2000}
\def \PDG{Particle Data Group, C. Caso \ite, \epjc{15}{1-878}{2000}}
\def \pisma#1#2#3#4{Pis'ma Zh.\ Eksp.\ Teor.\ Fiz.\ {\bf#1}, #2 (#3) [JETP
Lett.\ {\bf#1}, #4 (#3)]}
\def \pl#1#2#3{Phys.\ Lett.\ {\bf#1}, #2 (#3)}

\def \plb#1#2#3{Phys.\ Lett.\ B {\bf#1}, #2 (#3)}
\def \pr#1#2#3{Phys.\ Rev.\ {\bf#1}, #2 (#3)}

\def \prd#1#2#3{Phys.\ Rev.\ D {\bf#1}, #2 (#3)}
\def \prl#1#2#3{Phys.\ Rev.\ Lett.\ {\bf#1}, #2 (#3)}
\def \prp#1#2#3{Phys.\ Rep.\ {\bf#1}, #2 (#3)}
\def \ptp#1#2#3{Prog.\ Theor.\ Phys.\ {\bf#1}, #2 (#3)}
\def \rmp#1#2#3{Rev.\ Mod.\ Phys.\ {\bf#1}, #2 (#3)}

\def \si90{25th International Conference on High Energy Physics, Singapore,
Aug. 2-8, 1990}
\def \slc87{{\it Proceedings of the Salt Lake City Meeting} (Division of
Particles and Fields, American Physical Society, Salt Lake City, Utah, 1987),
ed. by C. DeTar and J. S. Ball (World Scientific, Singapore, 1987)}
\def \slac89{{\it Proceedings of the XIVth International Symposium on
Lepton and Photon Interactions,} Stanford, California, 1989, edited by M.
Riordan (World Scientific, Singapore, 1990)}
\def \smass82{{\it Proceedings of the 1982 DPF Summer Study on Elementary
Particle Physics and Future Facilities}, Snowmass, Colorado, edited by R.
Donaldson, R. Gustafson, and F. Paige (World Scientific, Singapore, 1982)}
\def \smass90{{\it Research Directions for the Decade} (Proceedings of the
1990 Summer Study on High Energy Physics, June 25--July 13, Snowmass, Colorado),
edited by E. L. Berger (World Scientific, Singapore, 1992)}
\def \tasi{{\it Testing the Standard Model} (Proceedings of the 1990
Theoretical Advanced Study Institute in Elementary Particle Physics, Boulder,
Colorado, 3--27 June, 1990), edited by M. Cveti\v{c} and P. Langacker
(World Scientific, Singapore, 1991)}

\def \zpc#1#2#3{Zeit.\ Phys.\ C {\bf#1}, #2 (#3)}

\begin{document}
\topmargin -0.1in
\title{CP Violation:  Past, Present, and Future$^*$}
\author{Jonathan L. Rosner}
\address{Enrico Fermi Institute and Department of Physics \\
University of Chicago, Chicago, IL 60637 USA}
\maketitle
\begin{abstract}
We discuss the history of CP violation and its
manifestations in kaon physics, its explanation in terms of phases of the
Cabibbo-Kobayashi-Maskawa matrix describing charge-changing weak quark
transitions, predictions for experiments involving
$B$ mesons, and the light it can shed on physics beyond the Standard Model.
\end{abstract}

\section{Introduction}

CP symmetry and its violation are important guides to fundamental quark
properties and to the understanding of the matter-antimatter asymmetry
of the Universe.  In this review, an updated version of one presented
earlier in the year \cite{PR}, we describe past, present, and future
aspects of CP violation studies.  After an illustration of fundamental
discrete symmetries in Maxwell's equations (Sec.\ II), we recall the history of
CP violation's discovery \cite{CCFT} in the decays of neutral kaons
(Sec.\ III).  The product CPT so far seems to be conserved, as is expected in
local Lorentz-invariant quantum field theories \cite{CPT}.  We then discuss the
electroweak theory's explanation of CP violation \cite{KM} in terms of phases
of the Cabibbo-Kobayashi-Maskawa (CKM) \cite{KM,Cab} matrix in
Sec.\ IV), and mention some present tests of this theory with kaons (Sec.\ V)
$B$ mesons (Sec.\ VI), and charmed particles (Sec.\ VII).  The future of CP
violation studies (Sec.\ VIII) is very rich, with a wide variety of experiments
relevant to physics beyond the Standard Model and the baryon asymmetry of
the Universe.

\section{DISCRETE SYMMETRIES IN MAXWELL'S EQUATIONS}

The behavior of the Maxwell equations under the discrete symmetries P
(parity), T (time reversal), C (charge conjugation), and CPT is summarized in
Table \ref{tab:max}.  Each term behaves as shown.

\renewcommand{\arraystretch}{1.4}
\begin{table}[h]
\caption{Behavior of Maxwell's equations under discrete symmetries.
\label{tab:max}}
\begin{center}
\begin{tabular}{c c c c c}
Equation & P & T & C & CPT \\
$\nabla \cdot {\bf E} = 4 \pi \rho$ & $+$ & $+$ & $-$ & $-$ \\
$\nabla \cdot {\bf B} = 0$          & $-$ & $-$ & $-$ & $-$ \\
$\nabla \times {\bf B} - \frac{1}{c} \frac{\partial {\bf E}}{\partial t} =
\frac{4 \pi}{c}{\bf j}$                 & $-$ & $-$ & $-$ & $-$ \\
$\nabla \times {\bf E} + \frac{1}{c} \frac{\partial {\bf B}}{\partial t} = 0$
                                    & $+$ & $+$ & $-$ & $-$ \\
\end{tabular}
\end{center}
\end{table}

Under P, we have
\bea
{\bf E}({\bf x},t) & \to & - {\bf E}(-{\bf x},t),~~~
{\bf B}({\bf x},t) \to {\bf B}(-{\bf x},t), \\
\nabla & \to & - \nabla,~~ {\bf j}({\bf x},t) \to - {\bf j}({\bf -x},t).
\eea
Electric fields change in sign while magnetic fields do not, and
currents change in direction.  Under T,
\bea
{\bf E}({\bf x},t) & \to & {\bf E}({\bf x},-t),~~~
{\bf B}({\bf x},t) \to - {\bf B}({\bf x},-t), \\
\partial/\partial t & \to & - \partial/\partial t,~~~
{\bf j}({\bf x},t) \to - {\bf j}({\bf x}, -t).
\eea
Magnetic fields change in sign while electric fields do not, since
directions of currents are reversed.  Under C,
\bea
{\bf E}({\bf x},t) & \to & - {\bf E}({\bf x},t),~~~
{\bf B}({\bf x},t) \to - {\bf B}({\bf x},t), \\
\rho({\bf x},t) & \to & - \rho({\bf x},t),~~~
{\bf j}({\bf x},t) \to - {\bf j}({\bf x}, t).
\eea
Both electric and magnetic fields change sign, since their sources
$\rho$ and ${\bf j}$ change sign.  Finally, under CPT, space and time are
inverted but electric and magnetic fields retain their signs:
\beq
{\bf E}({\bf x},t) \to {\bf E}({\bf -x},-t),~~~
{\bf B}({\bf x},t) = {\bf B}(-{\bf x},-t).
\eeq

A fundamental term in the Lagrangian
behaving as ${\bf E} \cdot {\bf B}$, while Lorentz covariant, would violate P
and T.  Such a term seems to be strongly suppressed, in view of the small
value of the neutron electric dipole moment.  Its absence is a mystery, but
several possible reasons have been proposed (see, e.g., \cite{SCPrev}).

\section{CP SYMMETRY FOR KAONS}

Some neutral particles, such as the photon, the neutral pion, and the $Z^0$,
are their own antiparticles, while some -- those carrying nonzero quantum
numbers -- are not.  The
neutral kaon $K^0$, discovered in 1946 \cite{RB} in cosmic radiation, was
assigned a ``strangeness'' quantum number $S=1$ in the classification
scheme of Gell-Mann and Nishijima \cite{GN} in order to explain its
strong production and weak decay.  Production would conserve strangeness,
while the weaker decay process would not.  For this scheme to make sense it
was then necessary that there also exist an anti-kaon, the $\ok$, with $S=-1$.

As Gell-Mann described this scheme at a seminar at the University of Chicago,
Enrico Fermi asked him what distinguished the $\ok$ from the $K^0$ if both
could decay to $\pi \pi$, as seemed to be observed.  This question led
Gell-Mann and Pais \cite{GP} to propose that the states of definite mass and
lifetime were

\bea
K_1 & = & \frac{K^0 + \ok}{\s}~~~(C = +), \\
K_2 & = & \frac{K^0 - \ok}{\s}~~~(C = -), \\
\eea
with the $K_1$ allowed by C invariance (then thought to be a property of
weak interactions) to decay to $\pi\pi$ and the $K_2$ forbidden to decay to
$\pi \pi$.  The $K_2$ would be allowed to decay only to three-body final states
such as $\pi^+ \pi^- \pi^0$ and thus would have a much longer lifetime.  It
was looked for and found in 1956 \cite{KL}.
The discovery that the weak interactions violated C and P but apparently
preserved the product CP \cite{CPK} led to a recasting of the above argument
through the identification $CP(K_1) = +(K_1)$, $CP(K_2) = -(K_2)$.

\begin{figure}
\centerline{\epsfysize = 2in \epsffile {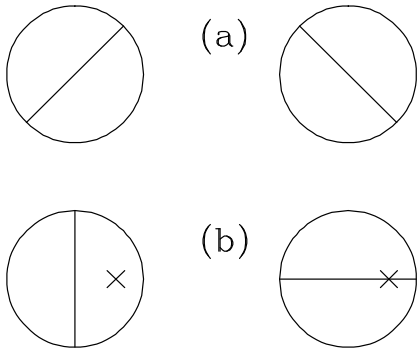}}
\caption{Basis states for first excitations of a drum head. (a) Nodal
lines at $\pm 45^\circ$ with respect to horizontal; (b) horizontal and
vertical nodal lines.
\label{fig:dh}}
\end{figure}

The $K_1$--$K_2$ system can be illustrated using a degenerate two-state
example such as a pair of coupled pendula \cite{BW} or
the first excitations of a drum head.  There is no way to
distinguish between the basis states illustrated in Fig.\ \ref{fig:dh}(a),
in which the nodal lines are at angles of $\pm 45^\circ$ with respect to
the horizontal, and those in Fig.\ \ref{fig:dh}(b), in which they are
horizontal and vertical. 

If a fly lands on the drum-head at the point marked ``$\times$'', the
basis (b) corresponds to eigenstates.  One of the modes couples to the
fly; the other doesn't.  The basis in (a) is like that of $(K^0,\ok)$, while
that in (b) is like that of $(K_1,K_2)$.  Neutral kaons are produced as in
(a), while they decay as in (b), with the fly analogous to the $\pi \pi$
state.  The short-lived state ($K_1$, in this CP-conserving approximation)
has a lifetime of 0.089 ns, while the long-lived state ($\simeq K_2$) lives
$\sim 600$ times as long, for 52 ns.

In 1964 Christenson, Cronin, Fitch, and Turlay \cite{CCFT} found that indeed
one in about 500 long-lived neutral kaons {\it did} decay to $\pi^+ \pi^-$, and
one in about 1000 decayed to $\pi^0 \pi^0$.  The states of definite mass and
lifetime could then be written, approximately, as
\bea
K_S~({\rm ``short"}) \simeq K_1 + \epsilon K_2, \nonumber \\
K_L~({\rm ``long"}) \simeq K_2 + \epsilon K_1 \label{eqn:es},
\eea
with a parameter $\epsilon$ whose magnitude was about $2 \times 10^{-3}$ and
whose phase was about $45^\circ$.  Since the states of definite mass and
lifetime were no longer CP eigenstates, CP had to be violated {\it somewhere}.
However, for many years $\epsilon$ was the only parameter describing CP
violation.  One could measure its magnitude and phase more and more precisely
(including learning about Re($\epsilon$) through a study of charge asymmetries
in $K_L \to \pi^\pm l^\mp \nu$), but its origin remained a mystery.  One
viable theory included a ``superweak'' one \cite{SW} which postulated a new
interaction mixing $K^0 = d \bar s$ and $\ok = s \bar d$ but with no other
consequences.

\begin{figure}
\centerline{\epsfysize = 1.3in \epsffile {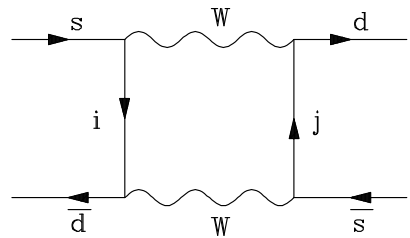}}
\caption{Box diagram describing the second-order weak mixing of a $K^0 =
d \bar s$ with a $\ok = s \bar d$.  There is another diagram with vertical
$W^+ W^-$ and horizontal quark-antiquark pairs $i,j = u,c,t$.
\label{fig:kbox}}
\end{figure}

Kobayashi and Maskawa offered a new opportunity to describe CP violation by
boldly postulating three quark families \cite{KM} when charm (the last member
of the second family) had not yet
even been firmly established.  In the diagram of Fig.\ \ref{fig:kbox}
describing the second-order weak transition $d \bar s \to s \bar d$ through
intermediate states involving pairs of quarks $i,j = u,c,t$ with charges 2/3,
the phases of complex weak couplings can have physical effects.  As long as
there are at least three quark families, one cannot redefine quark phases so
that all such couplings are real, and one can generate a nonzero value of
$\epsilon$.

The time-dependence of the two-component $\k$ and $\ok$ system is governed by
a $2 \times 2$ {\it mass matrix} $\m$ \cite{Revs}:
\beq
i \frac{\partial}{\partial t} \left[ \begin{array}{c} \k \\ \ok \end{array}
\right] = \m \left[ \begin{array}{c} \k \\ \ok \end{array} \right]~~~,
\eeq
where $\m = M - i \Gamma/2$, and $M$ and $\Gamma$ are Hermitian matrices.
The eigenstates (\ref{eqn:es}) then
correspond to the eigenvalues $\mu_{S,L} = m_{S,L} - i \gamma_{S,L}/2$, with
\beq
\epsilon \simeq \frac{{\rm Im}(\Gamma_{12}/2) + i~{\rm Im}~M_{12}}
{\mu_S - \mu_L}~~~.
\eeq
Using data and the magnitude of CKM matrix elements one can show
\cite{Revs} that the second term dominates.  Since the mass difference
$m_L - m_S$ and width difference $\gamma_S - \gamma_L$ are nearly equal,
the phase of $\mu_L - \mu_S$ is about $\pi/4$, so that the phase of $\epsilon$
is also $\pi/4$ (mod $\pi$).

\begin{figure}
\centerline{\epsfysize = 2in \epsffile {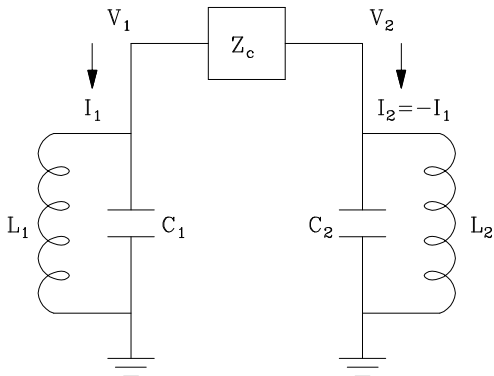}}
\caption{Coupled ``tank'' circuits illustrating the $K^0 - \bar K^0$ system.
The coupling impedance $Z_c$ must be asymmetric to emulate CP violation.
\label{fig:ckt}}
\end{figure}

It is easy to model the CP-conserving neutral kaon system in table-top
systems with two degenerate states \cite{BW}.  The
demonstration of CP violation requires systems that emulate Im($M_{12}) \ne 0$
or Im($\Gamma_{12}) \ne 0$.  One can couple two identical resonant circuits
``directionally'' to each other (see Fig.\ \ref{fig:ckt}) so that the energy
fed from circuit 1 to circuit 2 differs from that fed in the reverse direction
\cite{TTTV}.  Devices with this property utilize Faraday rotation of the plane
of polarization of radio-frequency waves; some references may be found in
\cite{Kost}.  This asymmetric coupling also is inherent in the
equations of motion of a spherical (or ``conical'') pendulum in a rotating
coordinate system \cite{RS}, so that the Foucault pendulum is a demonstration
(though perhaps not ``table-top'') of CP violation.  A ball rolling with
viscous damping in a rotating vase of elliptical cross section holds
more promise for a laboratory setting \cite{Kost}.  In all such
cases the CP-violating effect is imposed ``from the outside,'' leaving open
the question of whether some ``new physics'' is governing the corresponding
effect in particle physics.

\section{KOBAYASHI-MASKAWA THEORY OF CP VIOLATION}

The interactions of quarks with $W^\pm$ bosons are described by
\beq
{\cal L}_{\rm int} = \frac{g}{\s} [ \bar U'_L \gamma^\mu W_\mu^{(+)} D'_L
+ {\rm H.c.} ],
\eeq
where the primed quarks are ``weak eigenstates'':
\beq
U' \equiv \left[ \begin{array}{c} u' \\ c' \\ t' \end{array} \right],~~
D' \equiv \left[ \begin{array}{c} d' \\ s' \\ b' \end{array} \right].
\eeq
In the weak-eigenstate basis, the mass term in the Lagrangian,
\beq
{\cal L}_m = -[\bar U'_R {\m}_U U'_L + \bar D'_L {\m}_D D'_L + {\rm H.c.} ],
\eeq
will involve a general $3 \times 3$ matrix $\m$, which requires
separate left and right unitary transformations
\beq
R^{\dag}_Q {\m}_Q L_Q = \Lambda_Q
\eeq
to obtain a diagonal matrix $\Lambda_Q$ with non-negative entries.
If we define unprimed (mass) eigenstates by
\beq
Q'_L = L_Q Q_L,~~Q'_R = R_Q Q_R~~(Q=U,~D),
\eeq
the interaction Lagrangian may be expressed as
\beq
{\cal L}_{\rm int} = \frac{g}{\s} [\bar U_L \gamma^\mu W_\mu^{(+)} V
D_L + {\rm H.c.} ],
\eeq
where $V \equiv L^{\dag}_U L_D$ is the Cabibbo-Kobayashi-Maskawa (CKM)
matrix.  As a result of its unitarity, $V^\dag V = V V^\dag = 1$, the
$Z q \bar q$ couplings in the electroweak theory are flavor-diagonal.
Since it contains no information about $R_U$ or $R_D$,
$V$ provides only partial information about ${\m}_Q$.

For $n~u$-type quarks and $n~d$-type quarks, $V$ is $n \times n$.  Since it
is unitary, it can be described by $n$ real parameters.  Relative quark
phases account for $2n-1$ of these, leaving $n^2 - (2n-1) = (n-1)^2$ physical
parameters. Of these, $n (n - 1)/2$ (the number of independent rotations
in $n$ dimensions) correspond to angles, while the rest, $(n-1)(n-2)/2$,
correspond to phases.

\begin{figure}
\centerline{\epsfysize=2in \epsffile{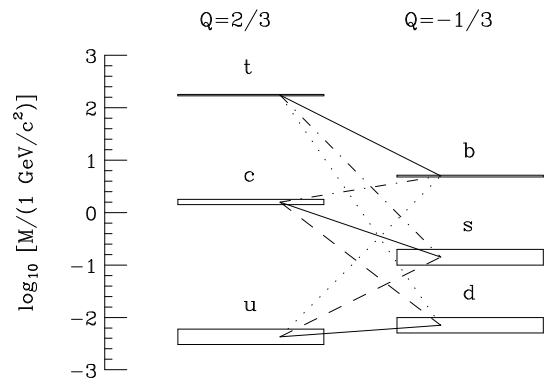}}
\caption{Pattern of charge-changing weak transitions among quarks.  Solid
lines:  relative strength 1; dashed lines:  relative strength 0.22;
dot-dashed lines:  relative strength 0.04; dotted lines:  relative strength
$\le 0.01$. Breadths of levels denote estimated errors in quark masses.
\label{fig:trans}}
\end{figure}

For $n=2$, we have one angle and no phases. The matrix $V$ then can always be
chosen as orthogonal \cite{Cab,Charm}.  For $n=3$, we have three angles and one
phase, which in general cannot be eliminated by arbitrary choices of phases in
the quark fields. It was this phase that motivated Kobayashi and Maskawa
\cite{KM} to introduce a third quark doublet in 1973 when only two were
known. (The bottom quark was discovered in 1977 \cite{ups}, and the top in 1994
\cite{top}.)  The Kobayashi-Maskawa theory provides a potential
source of CP violation, serving as the leading contender for the observed
CP-violating effects in the kaon system and suggesting substantial CP
asymmetries in the decays of mesons containing $b$ quarks.
The pattern of charge-changing weak transitions among quarks is depicted
in Fig.\ \ref{fig:trans}.

A convenient parametrization of the CKM matrix utilizes a hierarchy \cite{WP}
whereby magnitudes of elements are approximately powers of $\lambda \equiv
\sin \theta_c \simeq 0.22$, where
$\theta_c$ is the Gell-Mann--L\'evy--Cabibbo angle \cite{Cab,GL} describing 
strange particle decays.  The matrix may be expressed as
\beq
V = \left[ \begin{array}{c c c}
1 - \frac{\lambda^2}{2} & \lambda & A \lambda^3 (\rho - i \eta) \\
- \lambda & 1 - \frac{\lambda^2}{2} & A \lambda^2 \\
A \lambda^3(1 - \rho - i \eta) & - A \lambda^2 & 1 \end{array} \right],
\eeq
where rows denote $u,~c,~t$ and columns denote $d,~s,~b$.

We learn $|V_{cb}| = A \lambda^2 \simeq 0.041 \pm 0.003$ from the dominant
decays of $b$ quarks, which are to charmed quarks \cite{JRCKM,JRTASI}.
Smaller errors are quoted in most reviews \cite{CKMrevs} which
take different views of the dominantly theoretical sources of error.
As an indication that this number is still in some flux we note a new
measurement $|V_{cb}| = 0.046 \pm 0.004$ by the CLEO group \cite{CLEOVcb}.)
Similarly, we shall take from charmess $b$ decays $|V_{ub}/V_{cb}| = 0.090 \pm
0.025 = \lambda (\rho^2 + \eta^2)^{1/2}$ \cite{Flg}, leading to $\rho^2 +
\eta^2 = 0.41 \pm 0.11$, whereas smaller errors are quoted by most
authors.

\begin{figure}
\centerline{\epsfysize = 1.4in \epsffile {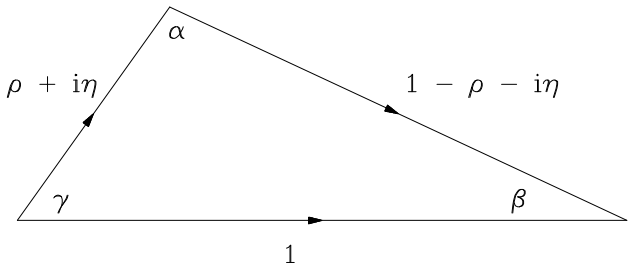}}
\caption{Unitarity triangle for CKM elements.  Here $\rho + i \eta =
V^*_{ub}/A \lambda^3$; $1 - \rho - i \eta = V_{td}/A \lambda^3$.
\label{fig:ut}} 
\end{figure}

As a result of the unitarity of the CKM matrix, the quantities $V^*_{ub}/A
\lambda^3 = \rho + i \eta$, $V_{td}/A \lambda^3 = 1 - \rho - i \eta$, and 1
form a triangle in the $(\rho,\eta)$ plane (Fig.\ \ref{fig:ut}).  We still do
not have satisfactory limits on the angle $\gamma$ of this ``unitarity
triangle.'' Further information comes from the following constraints:

1. {\it Mixing of neutral $B$ mesons} is
dominated by top quark contributions to graphs such as Fig.\ \ref{fig:kbox}
but with external quarks $d \bar b$ for $B^0$ or $s \bar b$ for $B_s$.
For example, the mass splitting in the nonstrange neutral $B$ system is
\beq
\Delta m_d = 0.487 \pm 0.014~{\rm ps}^{-1} \sim f_B^2 B_B |V_{td}|^2,
\eeq
where $f_B$ is the $B$ meson decay constant and $B_B = {\cal O}(1)$ is the
``vacuum saturation factor,'' describing the degree to which graphs such as
Fig.\ \ref{fig:kbox} describe the mixing.  Recent estimates \cite{lat} give
$f_B \sqrt{B_{B}} = 230 \pm 40$ MeV.  Consequently, one finds \cite{JRCKM}
$|1 - \rho - i \eta| = 0.87 \pm 0.21$.  Neutral strange $B$ mesons are
characterized by \cite{Bslim}
\beq
\Delta m_s \sim f_{B_s}^2 B_{B_s} |V_{ts}|^2 > 15~{\rm ps}^{-1}.
\eeq
Since $|V_{ts}| \simeq |V_{cb}$ is approximately known, this information
mainly serves to constrain the product $f_{B_s} \sqrt{B_{B_s}}$ and, given
information on the ratio of strange and nonstrange constants \cite{JRFM}, the
value of $|V_{td}|$, leading to $|1-\rho-i\eta| < 1.01$.  The large top mass,
$m_t = 174 \pm 5$ GeV \cite{PDG}, is crucial for these mixings to be so large.

2. {\it CP-violating $K^0$--$\ok$ mixing} through the box graphs
of Fig.\ \ref{fig:kbox} accounts for the parameter \cite{PDG}
\beq
\epsilon = (2.27 \times 10^{-3})e^{i 43.3^\circ} \sim {\rm Im}{\m}_{12}
\sim f_K^2 B_K~ {\rm Im}(V_{td}^2),
\eeq
leading to a constraint \cite{JRCKM,JRTASI}
\beq
\eta (1 - \rho + 0.39) = 0.35 \pm 0.12~~~.
\eeq
Here we have used $f_K = 161$ MeV and $B_K = 0.87 \pm 0.13$ \cite{Lubicz}.
If top quarks were fully dominant the left-hand side of this equation
would be just $\eta(1-\rho)$.  The term 0.39 in brackets is a correction due to
charmed quarks.

\begin{figure}
\centerline{\epsfysize = 1.8in \epsffile {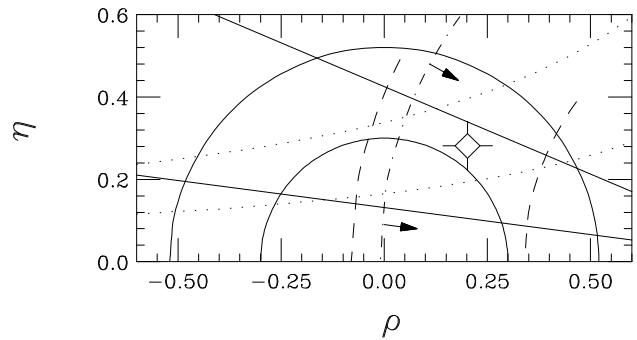}}
\caption{Region of $(\rho,\eta)$ specified by constraints on
CKM matrix parameters. Solid semicircles denote limits based on
$|V_{ub}/V_{cb}| = 0.090 \pm 0.025$; dashed arcs denote limits $0.66 \le
|1 - \rho - i \eta| \le 1.08$ based on $\b$--$\ob$ mixing; dot-dashed arc
denotes limit $|1 - \rho - i \eta| < 1.01$ based on $B_s$--$\overline{B_s}$
mixing; dotted lines denote limits $\eta (1 - \rho + 0.39) = 0.35 \pm 0.12$
based on CP-violating $\k$--$\ok$ mixing.  Rays:  $\pm 1 \sigma$ limits
on $\sin 2 \beta$ (see Sec.\ VI).  The plotted point at $(\rho,\eta)
\simeq (0.20,0.28)$ lies roughly in the middle of the allowed
region.
\label{fig:re}}
\end{figure}

The constraints are plotted on the $(\rho,\eta)$ plane in Fig.\ \ref{fig:re}.
Also shown are the $\pm 1 \sigma$ bounds on $\sin 2 \beta$, to be discussed
presently, from an average $0.49 \pm 0.23$ \cite{JRTASI} of OPAL, ALEPH, CDF,
BaBar, and BELLE values.  The allowed region is larger than that favored by
many other analyses \cite{CKMrevs}.

\section{THE CKM MATRIX AND PREDICTIONS FOR KAON PHYSICS}

\subsection{$K_{S,L} \to \pi \pi$ rates}

If we define
\beq
\eta_{+-} \equiv \frac{A(K_L \to \pi^+ \pi^-)}{A(K_S \to \pi^+ \pi^-)},~~
\eta_{00} \equiv \frac{A(K_L \to \pi^0 \pi^0)}{A(K_S \to \pi^0 \pi^0)},
\eeq
the possibility of different CP-violating effects in $\pi \pi$ states of
isospin $\ipp=2$ and $\ipp = 0$ \cite{paren} gives rise to a parameter
$\epsilon'$ such that $\eta_{+-} = \epsilon + \epsilon'$, $\eta_{00} =
\epsilon - 2 \epsilon'$.  The following ratio of ratios then can differ
from unity:
$$
R \equiv \frac{\Gamma(K_L \to \pi^+ \pi^-)}{\Gamma(K_S \to \pi^+ \pi^-)} /
\frac{\Gamma(K_L \to \pi^0 \pi^0)}{\Gamma(K_S \to \pi^0 \pi^0)}
$$
\beq
 = 1 + 6~{\rm Re}\frac{\epsilon'}{\epsilon}.
\eeq
The ratio $\epsilon'/\epsilon$ is expected to be approximately real in a
CPT-invariant theory \cite{Revs}.  A key prediction of the KM theory is that
$\epsilon'/\epsilon$ should be a number of order $10^{-3}$.  Two types of
amplitudes contribute to $K \to \pi \pi$ decays.

1. {\it Tree amplitudes}, involving the quark subprocess $s \to u \bar u
d$, have both $\Delta I=1/2$ and $\Delta I = 3/2$ components and thus
contribute to both $\ipp=0$ and $\ipp=2$ states.  In a standard 
convention \cite{WP}, tree amplitudes contain no weak phases, since they
involve the CKM elements $V_{ud}$ and $V_{us}$.

2. {\it Penguin amplitudes}, involving the quark subprocess $s \to d$
with an intermediate loop consisting of a $W$ boson and the quarks $u,~c,~t$,
and interacting with the rest of the system through one or more gluons,
have only $\Delta I=1/2$ and thus can only contribute to the $\ipp=0$
state.  The top quark in the loop gives rise to a weak phase through the
CKM element $V_{td}$. 

A relative weak phase of $\ipp=0$ and $\ipp=2$ states is thus generated in
the KM theory, leading to $\epsilon'/\epsilon \ne 0$. {\it Electroweak} penguin
amplitudes, in which the gluon connecting the $s \to d$ subprocess to the
rest of the diagram is replaced
by a photon or $Z^0$, can have both $\Delta I=1/2$ and $\Delta I = 3/2$
components and tend to reduce the predicted value of $\epsilon'/\epsilon$.
One range of estimates \cite{Buras} finds a broad and somewhat asymmetric
probability distribution extending from slightly below zero to above $2 \times
10^{-3}$.  Others (see articles in \cite{K99}) permit slightly higher values.

Recent experiments on Re($\epsilon'/\epsilon$) \cite{E731,NA31,E832,NA48} are 
summarized in Table \ref{tab:epe}. (The error in the average includes a scale
factor
\cite{PDG} of 1.86.) The magnitude of $\epsilon'/\epsilon$ is consistent with
estimates based on the Kobayashi-Maskawa theory.  The qualitative agreement is
satisfactory, given that we still cannot account reliably for the
large enhancement of $\Delta I = 1/2$ amplitudes with respect to $\Delta I =
3/2$ amplitudes in {\it CP-conserving} $K \to \pi \pi$ decays.  More data are
expected from the Fermilab and CERN experiments, reducing the eventual
statistical error on $\epsilon'/\epsilon$ to a part in $10^4$.

\begin{table}
\caption{Recent experimental values for Re$(\epsilon'/\epsilon)$.
\label{tab:epe}}
\begin{center}
\begin{tabular}{c c c c}
\protect
Experiment & Reference & Value ($\times 10^{-4}$) & $\Delta \chi^2$ \\ \hline
Fermilab E731 & \cite{E731} &  $7.4 \pm 5.9$ & 3.97 \\
CERN NA31     & \cite{NA31} & $23.0 \pm 6.5$ & 0.35 \\
Fermilab E832 & \cite{E832} & $28.0 \pm 4.1$ & 4.65 \\
CERN NA48     & \cite{NA48} & $14.0 \pm 4.3$ & 1.44 \\
Average       &             & $19.2 \pm 4.6$ & $\sum = 10.4$ \\
\end{tabular}
\end{center}
\end{table}

\subsection{$K \to \pi l^+ l^-$ information}

1.  {\it The decay $K^+ \to \pi^+ \nu \bar \nu$} involves loop diagrams
involving $V_{td}$ and a small charm correction in such a way that the
combination $|1.4 - \rho - i \eta|$ is constrained, with a predicted branching
ratio of order

\beq
{\cal B}(K^+ \to \pi^+ \nu \bar \nu) \simeq 10^{-10} \left|
\frac{|1.4 - \rho - i \eta}{1.4} \right|^2,
\eeq
or for the range permitted in Fig.\ \ref{fig:re}, a branching ratio of about
$(0.8 \pm 0.2) \times 10^{-10}$ \cite{BuK}.  Additional
uncertainties are associated with $m_c$ \cite{FalkK} and $|V_{cb}|$.  A
measurement of ${\cal B}(K^+ \to \pi^+ \nu \bar \nu)$ to 10\% will help to
constrain $(\rho,\eta)$ more tightly than in Fig.\ \ref{fig:re} or will
expose inconsistencies in our present picture of CP violation.

Up to now the Brookhaven E787 Collaboration sees only one $K^+ \to \pi^+ \nu 
\bar \nu$ event with negligible background \cite{E787}, corresponding to
\beq
{\cal B}(K^+ \to \pi^+ \nu \bar \nu) = (1.5^{+3.4}_{-1.2}) \times 10^{-10}~~~.
\eeq
More data are expected from the final analysis of this experiment,
as well as from a future version with improved sensitivity.

2.  {\it The decays $K_L \to \pi^0 l^+ l^-$} should be dominated by
CP-violating contributions, both indirect ($\sim \epsilon$) and direct, with a 
CP-conserving ``contaminant'' from $K_L \to \pi^0 \gamma \gamma \to \pi^0
l^+ l^-$. The direct contribution probes the parameter $\eta$.  Each
contribution (including the CP-conserving one) is expected to correspond to
a $\pi^0 e^+ e^-$ branching ratio of a few parts in $10^{12}$.
However, $K_L \to \pi^0 e^+ e^-$ may be limited by backgrounds
in the $\gamma \gamma e^+ e^-$ final state associated with radiation of a
photon in $K_L \to \gamma e^+ e^-$ from one of the leptons
\cite{HGr}.  Present experimental upper limits (90\% c.l.) \cite{pll} are
$$
{\cal B}(K_L \to \pi^0 e^+ e^-) < 5.1 \times 10^{-10},
$$
\beq
{\cal B}(K_L \to \pi^0 \mu^+ \mu^-) < 3.8 \times 10^{-10},
\eeq
still significantly above most theoretical expectations.  (See, however,
\cite{Ko}.)

3. {\it The decay $K_L \to \pi^0 \nu \bar \nu$} should be due entirely to
CP violation, and provides a clean probe of $\eta$.  Its branching
ratio, proportional to $A^4 \eta^2$, is expected to be about $3 \times
10^{-11}$.  The best current experimental upper limit (90\% c.l.) for this
process \cite{pnn} is ${\cal B}(K_L \to \pi^0 \nu \bar \nu) < 5.9 \times
10^{-7}$, several orders of magnitude above the expected value.

\subsection{Other rare kaon decays}

1.  {\it The decay $K_L \to \pi^+ \pi^- e^+ e^-$} involves three
independent momenta in the final state and thus offers the opportunity to
observe a T-odd observable through a characteristic distribution in the angle
$\phi$ between the $\pi^+ \pi^-$ and $e^+ e^-$ planes.  A CP- or T-violating
angular asymmetry in this process has recently been reported
\cite{KTeVa,NA48a}.

2.  {\it The decay $K_L \to \mu^+ \mu^- \gamma$} has been studied
with sufficiently high statistics to permit a greatly improved measurement
of the virtual-photon form factor in $K_L \to \gamma^* \gamma$ \cite{BQ}.
This measurement is useful in estimating the long-distance contribution to
the real part of the amplitude in $K_L \to \gamma^{(*)} \gamma^{(*)} \to
\mu^+ \mu^-$, which in turn allows one to limit the short-distance contribution
to $K_L \to \mu^+ \mu^-$.

\subsection{Is the CKM picture of CP violation correct?}

The KM theory is comfortable with the observed range of $\epsilon'/\epsilon$,
and its prediction for ${\cal B}(K^+ \to \pi^+ \nu \bar \nu)$ is consistent
with the one event seen so far.  Further anticipated tests are the measurement
$\eta$ through the decay $K_L \to \pi^0 \nu \bar \nu$ (see below), and the
search for CP violation in hyperon decays, which is already under way
\cite{CLhyp,Luk}.  One also looks forward to a rich set of effects in
decays of particles containing $b$ quarks, particularly $B$ mesons.  We now
describe the experiments and the effects they are expected to see.

\section{CP VIOLATION IN $B$ DECAYS}

\subsection{Current and planned experiments}

Asymmetric $e^+ e^-$ collisions are being studied at ``$B$ factories'':
the PEP-II machine at SLAC with the BaBar detector, and the KEK-B collider in
Japan with the Belle detector.  By July 2000, these detectors
had accumulated about 14 and 6 fb$^{-1}$ of data at the energy of the
$\Upsilon(4S)$ resonance, which decays almost exclusively to $B \bar B$
\cite{Bas2b,BEs2b}.  As of September, 2000, PEP-II and KEK-B were providing 
about 150 and 100 pb$^{-1}$ per day to their respective detectors.

Further data on $e^+ e^-$ collisions at the $\Upsilon(4S)$ will be provided by
the Cornell Electron Storage Ring with the upgraded CLEO-III detector.  The
HERA-b experiment at DESY in Hamburg hopes to study $b$ quark
production via the collisions of 920 GeV protons with a fixed target.  The
CDF and D0 detectors at Fermilab will devote a significant part of their
program at Run II of the Tevatron to $B$ physics.  One can
expect further results on $B$ physics from the general-purpose LHC detectors
ATLAS and CMS, and the dedicated detectors at LHC-b at CERN and BTeV at
Fermilab.

\subsection{Types of CP violation}

In contrast to neutral kaons, whose mass eigenstates differ in lifetime by
nearly a factor of 600, the corresponding $B^0$--$\ob$ mass eigenstates
are predicted to differ in lifetime by at most 10--20\% for strange $B$'s
\cite{BBD,bec}, and much less for nonstrange $B$'s.  Thus, instead of
mass eigenstates like $K_L$, two main types of $B$ decays are of interest:
decays to CP eigenstates, and ``self-tagging'' decays.  Both have their
advantages and disadvantages.

1.  {\it Decays to CP eigenstates $f = \pm {\rm CP}(f)$} utilize interference
between direct decays $\b \to f$ or $\ob \to f$ and the corresponding paths
involving mixing:  $\b \to \ob \to f$ or $\ob \to \b \to f$.  Final states
such as $f = J/\psi K_S$ provide examples in which one quark
subprocess is dominant.  In this case one measures $\sin 2 \beta$ with
negligible corrections.  For $f = \pi^+ \pi^-$, one would measure
$\sin 2 \alpha$ only if the direct decay were dominated by
a ``tree'' amplitude (the quark subprocess $b \to u \bar u d$).  With
contamination from the penguin subprocess $b \to d$ expected to be about 30\%
in amplitude, one must measure decays to other $\pi \pi$ states (such as
$\pi^\pm \pi^0$ and $\pi^0 \pi^0$) to sort out amplitudes
\cite{GrL}.  In decays to CP eigenstates, one must determine the flavor of
the decaying $B$ at time of production.

2.  {\it ``Self-tagging'' decays} involve final states $f$ such as $K^+
\pi^-$ which can be distinguished from their CP-conjugates $\bar f$.  A
CP-violating rate asymmetry arises when two weak amplitudes $a_i$ with weak
phases $\phi_i$ and strong phases $\delta_i$ ($i=1,2)$ interfere:
$$
A(B \to f) = a_1 e^{i(+\phi_1 + \delta_1)} + a_2 e^{i(+\phi_2 + \delta_2)}~~~,
$$
\beq
~~~~~A(\bar B \to \bar f) = a_1 e^{i(-\phi_1 + \delta_1)} + a_2 e^{i(-\phi_2
 + \delta_2)}~~~.
\eeq
The weak phase changes sign under CP-conjugation, while the strong
phase does not.  The rate asymmetry is then
$$
{\cal A}(f) \equiv \frac{\Gamma(f) - \Gamma(\bar f)}
{\Gamma(f) + \Gamma(\bar f)}
$$
\beq \label{eqn:as}
= \frac{2 a_1 a_2 \sin(\phi_1 - \phi_2) \sin(\delta_1 - \delta_2)}
{a_1^2 + a_2^2 + 2 a_1 a_2 \cos(\phi_1 - \phi_2) \cos (\delta_1 - \delta_2)}~~.
\eeq
The two amplitudes must have different weak {\it and} strong phases in order
for a rate asymmetry to be observable. The CKM theory predicts the weak phases,
but no reliable estimates of strong phases exist.  We shall note some ways
to avoid this problem.

\subsection{Decays to CP eigenstates}

The interference between direct and mixing terms in $B$ decays to CP
eigenstates modulates the exponential decay (see, e.g., \cite{DR}):
\beq
\frac{d \Gamma(t)}{d t} \sim e^{- \Gamma t} (1 \mp {\rm Im} \lambda_0 \sin
\Delta m t),
\eeq
where the upper sign refers to $\b$ decays and the lower to $\ob$ decays.
$\Delta m$ is the mass splitting, and
$\lambda_0$ expresses the interference of decay and mixing amplitudes.  For
$f = J/\psi K_S$, $\lambda_0 = -e^{-2 i \beta}$, while
for $f = \pi^+ \pi^-$, $\lambda_0 \simeq e^{2 i \alpha}$ only to the extent
that penguin amplitudes can be neglected in comparison with
the dominant tree contribution.  The time integral of the modulation term is
\beq
\int_0^\infty dt e^{- \Gamma t} \sin \Delta m t = \frac{1}{\Gamma} \frac{x}
{1 + x^2} \le \frac{1}{\Gamma} \cdot \frac{1}{2}~~~,
\eeq
where $x \equiv \Delta m/\Gamma$.  This expression is maximum for $x = 1$,
and 96\% of maximum for the observed value $x \simeq 0.76$.

The CDF Collaboration \cite{CDFs2b} ``tags'' neutral $B$
mesons at the time of their production and measures the decay
rate asymmetry in $\b~(\ob) \to J/\psi K_S$.  This asymmetry arises from the
phase $2 \beta$ characterizing the two powers of $V_{td}$ in the $\b$--$\ob$
mixing amplitude. The tagging methods are of two main types. In
``opposite-side'' methods, since strong interactions produce $b$ and $\bar
b$ in pairs, one learns the initial flavor of a decaying $B$ from the ``other''
$b$-containing hadron produced in association with it.  ``Same-side''
methods \cite{GNR} utilize the fact that a $\b$ tends to be associated more
frequently with a $\pi^+$, and a $\ob$ with a $\pi^-$, somewhere nearby in
phase space.

Electron-positron collisions provide $B$ mesons in pairs at the
c.m. energy of the $\Upsilon(4S)$ resonance, just above threshold,
in states of negative charge-conjugation eigenvalue.  It then
becomes necessary to distinguish the vertices of the decaying and tagging $B$'s
from one another when studying CP eigenstates.  If $t$ and $t'$ denote the
decay and tagging proper times, the asymmetry for decay to a CP eigenstate will
be proportional to $\sin \Delta m (t-t')$, which vanishes when integrated over
all times (see, e.g., \cite{JRTASI} or \cite{BaBarbk}). The BaBar and BELLE
results were obtained using asymmetric $e^+ e^-$ collisions,
with typical vertex separations of about 250
$\mu$m and 200 $\mu$m.  PEP-II, constructed in the ring of the old PEP machine,
collides 9 GeV electrons with 2.7 GeV positrons, while KEK-B, constructed in
the TRISTAN tunnel, collides 8.5 GeV electrons with 3.5 GeV positrons.  In
symmetric collisions the $\Upsilon(4S)$ is produced at rest and the proper
path length of a decaying $B$ is only about 30 $\mu$m.

\begin{table}
\caption{Samples reported in July 2000 by BaBar and BELLE Collaborations
relevant to measurement of $\sin 2 \beta$. \label{tab:samp}}
\begin{tabular}{c c c c}
Collab. & Final state & Number & No.\ tagged \\ \hline
BaBar & $J/\psi K_S \to J/\psi \pi^+ \pi^-$ & 121 & 85 (50 $B^0$, 35 $\ob$) \\
      & $J/\psi K_S \to J/\psi \pi^0 \pi^0$ &  19 & 12 ( 7 $B^0$,  5 $\ob$) \\
      & $\psi'  K_S \to J/\psi \pi^+ \pi^-$ &  28 & 23 (13 $B^0$, 10 $\ob$) \\
      & Total                               & 168 & 120 \\
\hline
BELLE & CP-odd modes                        &  92 & 52 (40 $J/\psi K_S$) \\
      & $J/\psi K_L$                        & 102 & 42 \\
      & $J/\psi \pi^0$                      &  10 &  4 \\
      & Total                               & 204 & 98 \\
\end{tabular}
\end{table}

Both BaBar and BELLE used tags based on leptons and kaons from $B$ decays.
BaBar also used two neural net methods.  The samples reported by the summer
of 2000 \cite{Bas2b,BEs2b} are shown in Table \ref{tab:samp}.

The CDF result and ones from OPAL \cite{OPs2b} and ALEPH \cite{ALs2b} utilizing
$B$'s produced in the decays of the $Z^0$ are compared with those from BaBar
and BELLE in Table \ref{tab:s2b}.  The average \cite{JRTASI} corresponds to the
$\pm 1 \sigma$ rays plotted in Fig.\ \ref{fig:re}.  There is no contradiction
(yet!) with the allowed region, but we look forward eagerly to reduced errors
from BaBar and BELLE.  New results are due to be presented in February of
2001.

\subsection{``Self-tagging'' decays}

\begin{figure}
\centerline{\epsfysize=1.1in \epsffile{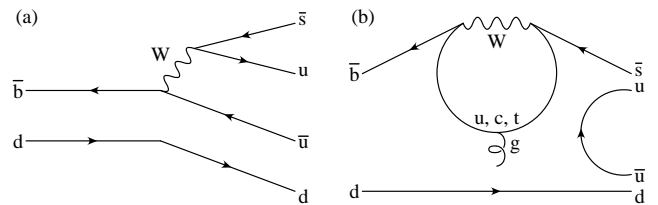}}
\caption{Contributions to $\b \to K^+ \pi^-$.  (a) Color-favored
``tree'' amplitude
$\sim V^*_{ub}V_{us}$; (b) ``penguin'' amplitude $\sim V^*_{tb}V_{ts}$.}
\label{fig:tp}
\end{figure}

A typical ``self-tagging'' mode suitable for the study of ``direct'' CP
violation is $B^0 \to K^+ \pi^-$.  The tree amplitude [Fig.\ \ref{fig:tp}(a)]
involves the quark subprocess $\bar b \to \bar s u \bar u$ with CKM factor
$V^*_{ub} V_{us}$ (weak phase $\gamma$).  The penguin amplitude [Fig.\
\ref{fig:tp}(b)] $\bar b \to \bar s$ with intermediate
$u,~c,~t$ quarks has CKM factor $V^*_{tb} V_{ts}$ or $V^*_{cb} V_{cs}$ (weak
phase $\pi$ or 0), depending on how the unitarity of the CKM matrix is used.  
The relative weak phase between the tree and penguin amplitudes thus is
non-zero, and direct CP violation can arise if the relative strong phase
$\delta_T - \delta_P$ also is non-zero.  The interpretation of a rate
difference $\Gamma(B^0 \to K^+ \pi^-) \ne \Gamma(\ob \to K^- \pi^+)$
requires independent information on $\delta_T - \delta_P$.

\begin{table}
\caption{Values of $\sin 2 \beta$ implied by recent measurements of the
CP-violating asymmetry in $B^0 \to J/\psi K_S$. \label{tab:s2b}}
\begin{center}
\begin{tabular}{c c}
\protect
Experiment & Value \\ \hline
OPAL \cite{OPs2b}  & $3.2^{+1.8}_{-2.0} \pm 0.5$ \\
CDF \cite{CDFs2b}  & $0.79^{+0.41}_{-0.44}$ \\
ALEPH \cite{ALs2b} & $0.84^{+0.82}_{-1.04} \pm 0.16$ \\
BaBar \cite{Bas2b} & $0.12 \pm 0.37 \pm 0.09$ \\
BELLE \cite{BEs2b} & $0.45^{+0.43+0.07}_{-0.44-0.09}$ \\ \hline
Average & $0.49 \pm 0.23$ \\
\end{tabular}
\end{center}
\end{table}

If one measures both a CP-violating asymmetry and a rate ratio such as
$\Gamma(B \to K^\pm \pi^\mp)/\Gamma(B^\pm \to K \pi^\pm)$ or
$\Gamma(B^\pm \to K^\pm \pi^0)/\Gamma(B^\pm \to K \pi^\pm)$, one can
eliminate the strong phase difference and solve for $\gamma$ \cite{GR,FM,NR}.
One must deal with electroweak penguins (which also affected the interpretation
of $\epsilon'/\epsilon$).  One proposal (see the first of Refs.~\cite{GR}) to
extract $\gamma$ from the rates for $B^+ \to (\pi^0 K^+, \pi^+ K^0,
\pi^+ \pi^0)$ and the charge-conjugate processes was flawed by the neglect
of these contributions, which are important \cite{DH}.  However, they
can be calculated \cite{NR}, so that measurements of the rates for these
processes can yield useful information on $\gamma$.

A necessary condition for the observability of direct CP asymmetries based on
the interference of two amplitudes, one weaker than the other, is that one must
be able to detect processes at the level of the absolute square of the weaker
amplitude \cite{EGR}.  Let the weak phase difference $\Delta \phi$ and the
strong phase difference $\Delta \delta$ both be near $\pm \pi/2$ (the most
favorable case).  Then the rate asymmetry ${\cal A}$ in Eq.\ (\ref{eqn:as}) has
magnitude
\beq
|{\cal A}| = {\cal O} \left( \frac{2 A_1 A_2}{A_1^2 + A_2^2} \right) \simeq
\frac{2 A_2}{A_1}~~~{\rm for}~A_2 \ll A_1.
\eeq
Define a rate based on the square of each amplitude:  $N_i = {\rm const.}~
|A_i|^2$.  Then $|{\cal A}| \simeq 2 \sqrt{N_2/N_1}$.

The statistical error in ${\cal A}$ is based on the total number of events.
For $A_2 \ll A_1$, one has $\delta {\cal A} \simeq 1/\sqrt{N_1}$.  Then the
significance of the asymmetry (in number of standard deviations) is
\beq
\left| \frac{{\cal A}}{\delta{\cal A}} \right| \sim {\cal O}(2 \sqrt{N_2}).
\eeq
Thus (aside from the factor of 2) one must be able to see the {\it square of
the weaker amplitude} at a significant level in order to see a significant
asymmetry due to $A_1$--$A_2$ interference.

In searching for direct CP asymmetries one thus considers $B$ decays with
at least two amplitudes having an expected weak phase difference, with a
large enough rate that the smaller amplitude alone would be detectable, and
with a good chance for a strong phase difference.

Many branching ratios for charmless $B$ decays are one to several
parts in $10^5$.  Rates associated with the subdominant amplitudes are expected
to be $\lambda^2 \simeq 1/20$ of these.  Thus when sensitivities to branching
ratios of a few parts in $10^7$ are reached, searches for direct CP asymmetries
will take on great significance.

Two processes whose rates favor a weak phase $\gamma$ exceeding $90^\circ$ are
$\b \to \pi^+ \pi^-$ and $\b \to K^{*+} \pi^-$ \cite{NR,GRg,Hou}, which favor
destructive and constructive tree-penguin interference, respectively. A  fit to
these and other processes in the second of Refs.\ \cite{Hou} finds $\gamma =
(114^{+24}_{-23})^\circ$, grazing the allowed region of Fig.\ \ref{fig:re}
but inconsistent with some more restrictive fits \cite{CKMrevs}.  Since the
upper bound on $\gamma$ is set by the limit on $B_s$--$\overline{B_s}$ mixing,
$\Delta m_s > 15$ ps$^{-1}$, such mixing should be visible soon.  There
is a hint of a signal at $\sim 17$ ps$^{-1}$ \cite{Bslim}.

The Tevatron and the LHC will produce many neutral $B$'s decaying to $\pi^+
\pi^-$, $K^\pm \pi^\mp$, and $K^+ K^-$ \cite{WurtJesik}.
Each of these channels has particular advantages.

1.  {\it The decays $\b \to K^+ K^-$ and $B_s \to \pi^+ \pi^-$} should be
suppressed unless these final states are ``fed'' by rescattering from other
channels \cite{resc}.

2.  {\it The decays $\b \to \pi^+ \pi^-$ and $B_s \to K^+ K^-$} can yield
$\gamma$ via time-dependence measurements \cite{RFKK}.

3.  {\it A recent proposal for measuring $\gamma$} \cite{bskpi} utilizes the
decays $\b \to K^+ \pi^-$, $B^+ \to \k \pi^+$, $B_s \to K^- \pi^+$, and the
corresponding charge-conjugate processes.  The $\b \to K^+ \pi^-$ and
$B_s \to K^- \pi^+$ peaks are well separated from one another and from
$\b \to \pi^+ \pi^-$ and $B_s \to K^+ K^-$ kinematically \cite{WurtJesik}.

The proposal of Ref.~\cite{bskpi} is based on the observation that $B \to K
\pi$ decays involve tree ($T$) and penguin ($P$) amplitudes
with relative weak phase $\gamma$ and relative strong phase $\delta$.
The decays $B^\pm \to K \pi^\pm$ are expected to be dominated by the penguin
amplitude (there is no tree contribution except through rescattering from other
final states), so this channel is not expected to display any CP-violating
asymmetries.  The prediction $\Gamma(B^+ \to \k \pi^+) = \Gamma(B^- \to \ok
\pi^-)$ thus will check the assumption that rescattering
effects can be neglected.  A typical amplitude is given by $A(\b \to K^+
\pi^-) = - [P + T e^{i(\gamma + \delta)}]$, where the signs are associated
with phase conventions for states \cite{GHLR}.  Defining
\beq
\left\{ \begin{array}{c} R \\ A_0 \end{array} \right\}
\equiv \frac{\Gamma(\b \to K^+ \pi^-)
\pm \Gamma(\ob \to K^- \pi^+)}{2 \Gamma(B^+ \to \k \pi^+)},
\eeq
\beq
\left\{ \begin{array}{c} R_s \\ A_s \end{array} \right\}
\equiv \frac{\Gamma(B_s \to K^- \pi^+)
\pm \Gamma(\overline{B_s} \to K^+ \pi^-)}{2 \Gamma(B^+ \to \k \pi^+)},
\eeq
and $r \equiv T/P$, $\tl \equiv V_{us}/V_{ud}$, one finds
$$
R = 1 + r^2 + 2 r \cos \delta \cos \gamma,
$$
\beq
R_s = \tl^2 + (r/\tl)^2 - 2 r \cos \delta \cos \gamma,
\eeq
\beq
A_0 = - A_s = -2 r \sin \gamma \sin \delta.
\eeq
The sum of $R$ and $R_s$ allows one to determine $r$.  Using $R$, $r$, and
$A_0$, one can solve for both $\delta$ and $\gamma$.  The prediction $A_s =
- A_0$ checks the flavor SU(3) assumption on which these relations are based.
An error of $10^\circ$ on $\gamma$ seems feasible with forthcoming
Tevatron data.

Recent upper limits on CP-violating asymmetries in $B$ decays to light-quark
systems \cite{CLEOCP}, defined as
\beq
{\cal A}_{CP} \equiv
 \frac{\Gamma(\overline{B} \to \bar f) - \Gamma(B \to f)}
{\Gamma(\overline{B} \to \bar f) + \Gamma(B \to f)},
\eeq
are shown in Table \ref{tab:CPA}.  No significant asymmetries have been seen,
but sensitivities adequate to check the maximum predicted values \cite{comb}
$|{\cal A}_{CP}^{K^+ \pi}| \le 1/3$ are being approached.

\begin{table}
\caption{CP-violating asymmetries in decays of $B$ mesons to light quarks.
\label{tab:CPA}}
\begin{center}
\begin{tabular}{c c c}
Mode & Signal events & ${\cal A}_{CP}$ \\ \hline
$K^+ \pi^-$    & $80^{+12}_{-11}$      & $-0.04 \pm 0.16$ \\
$K^+ \pi^0$    & $42.1^{+10.9}_{-9.9}$ & $-0.29 \pm 0.23$ \\
$K_S \pi^+$    & $25.2^{+6.4}_{-5.6}$  & $+0.18 \pm 0.24$ \\
$K^+ \eta'$    & $100^{+13}_{-12}$     & $+0.03 \pm 0.12$ \\
$\omega \pi^+$ & $28.5^{+8.2}_{-7.3}$  & $-0.34 \pm 0.25$ \\
\end{tabular}
\end{center}
\end{table}

\section{THE ROLE OF CHARM}

\subsection{Mixing and CP violation}

The dominant decay modes of the neutral charmed mesons $\d$ and $\od$ are
to states of negative and positive strangeness, respectively, and not to CP
eigenstates.  Thus $\d$--$\od$ mixing induced by shared final states
is expected to be small.  Short-distance contributions to mixing also are 
expected to be small.  Thus, in contrast to the case of neutral kaons and $B$
mesons, one expects small mass splittings, $\Delta m/\Gamma \ll 1$, and, in
contrast to neutral kaons, also small width differences.  The degree to which
cancellations among contributions of intermediate states such as $\pi^+ \pi^-$,
$K^+ K^-$, and $K^\pm \pi^\mp$ to mixing suppress such effects further is a
matter of debate \cite{HG}.  If
any rate difference is expected, it would be in the direction favoring a
slightly greater rate for the CP-even mass eigenstate.

CP violation in the charm sector is expected to be small in the Standard
Model.  It is also easy to look for, since $D$ mesons are easier to
produce than $B$ mesons and the Standard Model background is low.

Recent interesting studies of mixing by the CLEO \cite{CLEOmix} and
FOCUS \cite{FOCUSmix} Collaborations hint at the possibility of non-zero
values of $\Delta m$, $\Delta \Gamma$, or both, but are not yet
statistically compelling.  No evidence for mixing is found by the Fermilab
E791 Collaboration \cite{E791}. It may be necessary to invoke large final-state
phase differences in order to reconcile the CLEO and FOCUS results
\cite{BGLNP}.  No CP-violating asymmetries have been seen in
charmed meson decays at the level of several percent \cite{E791,CPcharm}.

\subsection{Spectroscopy}

A wide variety of excited $cqq$ and $c \bar q$ states are accessible at
CLEO and FOCUS.  The $cqq$ states are providing unique insights into baryon
spectroscopy \cite{CLEOomc,CLEOlc,CLEOxic}, while the $c \bar q$ states
\cite{CLEOD**,FOCUSspec}, are important sources of incormation about the
corresponding $b \bar q$ states, useful for ``same-side'' tagging of neutral
$B$ mesons.

\section{THE FUTURE}

\subsection{Envisioned measurements}

Future CP studies involve a broad program of experiments with kaons, charmed
and $B$ mesons, and neutrinos.

1.  {\it Rare kaon decays:}
Measurement of the branching ratio for $K_L \to \pi^0 \nu \bar \nu$ at the
required sensitivity (${\cal B} \simeq 3 \times 10^{-3}$) is foreseen at
Brookhaven National Laboratory \cite{K0pio} and the Fermilab Main Injector
\cite{KAMI}.  A Fermilab proposal \cite{CKM} seeks to acquire enough events of
$K^+ \to \pi^+ \nu \bar \nu$ to measure $|V_{td}|$ to a precision of 10\%.

2.  {\it Charmed mesons:}
While great strides have been taken in the measurement of mass and
lifetime differences for CP eigenstates of the neutral charmed mesons $D^0$,
\cite{CLEOmix,FOCUSmix}, it would be worth while to follow up present hints
of nonzero effects.  Both electron-positron colliders and hadronic experiments
devoted to future $B$ studies may also have more to say about mixing,
lifetime differences, and CP violation for charmed mesons.

3.  {\it $B$ production in symmetric $e^+ e^-$ collisions:}
Although asymmetric $e^+ e^-$ colliders are now taking data at an impressive
rate, the CLEO Collaboration is continuing with an active program.  It will be
able to probe charmless $B$ decays down to branching ratios of
$10^{-6}$.  It may be able to detect the elusive $\b \to \pi^0
\pi^0$ mode, whose rate will help pin down the penguin amplitude's
contribution and permit a determination of the CKM phase $\alpha$ \cite{GrL}.
Other final states of great interest at this level include $VP$ and $VV$,
where $P,V$ denote light pseudoscalar and vector mesons.
A useful probe of rescattering effects \cite{resc} is
the decay $\b \to K^+ K^-$.  This decay is expected to have a branching
ratio of only a few parts in $10^8$ if rescattering is unimportant, but could
be enhanced by more than an order of magnitude in the presence of rescattering
from other channels.  A challenging but crucial channel is $B^+ \to \tau^+ \bar
\nu_\tau$, whose rate will provide information on the combination $f_B
|V_{cb}|$.  Rare decays such as $B \to X \ell^+ \ell^-$ and $B \to X \nu \bar
\nu$ will probe the effects of new particles in loops.

4.  {\it $B$ production in asymmetric $e^+ e^-$ collisions:}
The BaBar and Belle detectors have made a start at the measurement
of $\sin 2 \beta$ in $\b \to J/\psi K_S$.  The moving center-of-mass
facilitates both flavor
tagging and improvement of signal with respect to background.  These machines
will make possible a host of time-dependent studies in such decays as $B \to
\pi \pi$, $B \to K \pi$, etc., and their impressive luminosities will
eventually add significantly to the world's tally of detected $B$'s.

5.  {\it Hadronic $B$ production:}
The strange $B$'s cannot be produced at the $\Upsilon(4S)$ which will dominate
the attention of $e^+ e^-$ colliders for some years to come. Hadronic reactions
at high energies will produce copious $b$'s incorporated into nonstrange,
strange, and charmed mesons, and baryons.  A measurement of the strange-$B$
mixing parameter $\Delta m_s$is likely to be made soon.  $B_s$ decays provide
valuable information on CKM phases and CP violation, as in $B_s \to K^+ K^-$
\cite{RFKK}.  The width difference expected between the CP-even and CP-odd
eigenstates of the $B_s$ system \cite{BBD,bec}
should be visible in the next round of experiments.

6.  {\it Neutrino studies:}
The magnitudes and phases in the CKM matrix are connected with the quark masses
themselves, whose pattern we will not understand until we have mapped out a
similar pattern for the leptons.  We will learn much about neutrino masses and
mixings from forthcoming experiments at the Sudbury Neutrino
Observatory \cite{SNO}, Borexino \cite{Bxo}, K2K \cite{Kam}, and Fermilab
(BooNE and MINOS) \cite{Fnu}.

\subsection{A likely parameter space}

\begin{figure}
\centerline{\epsfysize=2in \epsffile{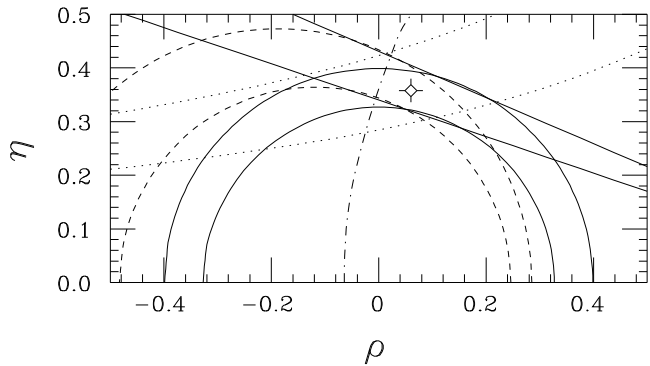}}
\caption{Plot in $(\rho,\eta)$ of anticipated constraints on CKM
parameters in the year 2003.  Solid curves: $|V_{ub}/V_{cb}|$; dashed lines:
constraint on $|V_{ub}/V_{td}|$ by combining measurement of ${\cal B}(B^+ \to
\tau^+ \nu_\tau)$ with $\b$--$\ob$ mixing; dotted lines: constraint due to
$\epsilon_K$ (CP-violating $\k$--$\ok$ mixing); dash-dotted line: limit
due to $x_s$; solid rays:  measurement of $\sin 2 \beta$ to $\pm 0.06$.
\label{fig:fut}}
\end{figure}

Our knowledge of the Cabibbo-Kobayashi-Maskawa is likely to improve
over the next few years \cite{JRlat,fut}.  With $\sin(2 \beta)$
measured in $\b \to J/\psi K_S$ decays to an accuracy of $\pm 0.06$ (the
BaBar goal with 30 fb$^{-1}$ \cite{BaBarbk}), errors on $|V_{ub}/V_{cb}|$
reduced to 10\%, strange-$B$ mixing bounded by $x_s = \Delta m_s/\Gamma_s
> 20$ (the present bound is already better than this!), and ${\cal B}(B^+ \to
\tau^+ \nu_\tau)$ measured to $\pm 20\%$ (giving $f_B|V_{ub}|$, or $|V_{ub}/\
V_{td}|$ when combined with $\b$--$\ob$ mixing), one finds the result
shown in Fig.\ \ref{fig:fut}.

The narrow range of $(\rho,\eta)$ increases the chance that any non-standard
physics will show up as a contradiction among various measurements, most
likely by providing additional contributions to $\b$--$\ob$ mixing \cite{GLmix}
but possibly directly affecting decays \cite{GW}.

\subsection{Baryon number of the Universe}

The number of baryons in the Universe is much larger than the corresponding
number of antibaryons.  Sakharov proposed \cite{Sakh} three requirements for
this preponderance of matter over antimatter:  (1) an epoch in which
the Universe was not in thermal equilibrium, (2) an interaction violating
baryon number, and (3) CP (and C) violation.  The observed baryon asymmetry is
not explained directly by the CP violation in the CKM matrix; the effects are
too small, requiring some new physics.  Two examples are the following:

1.  {\it Supersymmetry}, in which each particle of spin $J$ has
a ``superpartner'' of spin $J \pm 1/2$, affords many opportunities for
introducing new CP-violating phases and interactions which could affect
particle-antiparticle mixing \cite{SSBrev}.

2.  {\it Neutrino masses at the sub-eV level} can signal large right-handed
neutrino Majorana masses, exceeding $10^{11}$ GeV \cite{Ram}.
Lepton number ($L$), violated by such masses, can
be reprocessed into baryon number ($B$) by $B-L$ conserving
interactions at the electroweak scale \cite{LepBar}.  New CP-violating
interactions must exist at the high mass scale if lepton number is to
be generated there.  These interactions could be related to CKM phases
\cite{DPF}.  If this alternative is correct, it will be
important to understand the leptonic analogue of the CKM matrix!

\subsection{Surprises ahead?}

The CKM theory of CP violation in neutral kaon decays has passed a crucial
test.  The parameter $\epsilon'/\epsilon$ is nonzero, and has the expected
order of magnitude.  Tests using $B$ mesons, including the observation of a
difference in rates between $\b \to J/\psi K_S$ and $\ob \to J/\psi K_S$,
are just around the corner.  Progress in ``tagging'' neutral $B$'s and
rich information from measurements of many $B$ decay rates will round out
the picture.

If $B$ decays do not provide a consistent set of CKM phases in the next few
years, we will re-examine other proposed sources of CP violation.
Most of these, in contrast to the CKM theory, predict neutron
and electron dipole moments very close to their present experimental upper
limits. If, however, the CKM picture remains self-consistent, we should ask
about the origin of the CKM phases and the associated quark and
lepton masses.  It is probably time to start anticipating this possibility,
given the resilience of the CKM picture since it was first proposed nearly
30 years ago.

I am looking forward to a surprise such as one encountered
many years ago when exploring a small cave in Pennsylvania.  We had entered
it in the afternoon and thought we had seen all its rooms, when I came upon
another chamber with ghostly stalactites silhouetted against the
darkness behind them.  A breeze of warm air signaled that I was actually
looking outside, with the ``stalactites'' the faintly
glowing night sky, and the dark spaces the shadows of pine trees.  Such a
``perception shift'' does not come often, but is a welcome source of wonder.

\section{ACKNOWLEDGMENTS}

It is a pleasure to thank Adriano Natale and Rogerio Rosenfeld for the
invitation to attend the Brazilian Meeting of Particles and Fields and
for their wonderful hospitality in Sao Paulo and Sao Louren\c{c}o, and
Patrick Roudeau for constructive correspondence.
This work was supported in part by the United States
Department of Energy under Grant No.\ DE FG02 90ER40560.

\end{document}